\documentclass[aps,prb,jmp,amsmath,amssymb,reprint,twocolumn,superscriptaddress
]{revtex4-1}

\usepackage[english]{babel}

\usepackage{graphicx}
\usepackage{dcolumn}
\usepackage{bm,color}
\usepackage{epstopdf}

\begin{document}

\title{Second harmonic generation from strongly coupled localized and propagating phonon-polariton modes}

\date{\today}

\author{Ilya Razdolski}\email{razdolski@fhi-berlin.mpg.de}
\affiliation{Fritz Haber Institute of the Max Planck Society, 14195 Berlin, Germany}
\author{Nikolai Christian Passler}
\affiliation{Fritz Haber Institute of the Max Planck Society, 14195 Berlin, Germany}
\author{Christopher R. Gubbin}
\affiliation{School of Physics and Astronomy, University of Southampton, Southampton, SO17 1BJ, United Kingdom}
\author{Christopher J. Winta}
\affiliation{Fritz Haber Institute of the Max Planck Society, 14195 Berlin, Germany}
\author{Robert Cernansky}
\affiliation{School of Physics and Astronomy, University of Southampton, Southampton, SO17 1BJ, United Kingdom}
\author{Francesco Martini}
\affiliation{School of Physics and Astronomy, University of Southampton, Southampton, SO17 1BJ, United Kingdom}
\author{Alberto Politi}
\affiliation{School of Physics and Astronomy, University of Southampton, Southampton, SO17 1BJ, United Kingdom}
\author{Stefan A. Maier}
\affiliation{Chair in Hybrid Nanosystems, Nanoinstitute Munich, Faculty of Physics, Ludwig-Maximilians-Universit{\"a}t M{\"u}nchen, M{\"u}nchen, Germany}
\affiliation{Department of Physics, Blackett Laboratory, Imperial College London, London SW7 2AZ, United Kingdom}
\author{Martin Wolf}
\affiliation{Fritz Haber Institute of the Max Planck Society, 14195 Berlin, Germany}
\author{Alexander Paarmann}
\affiliation{Fritz Haber Institute of the Max Planck Society, 14195 Berlin, Germany}
\author{Simone De Liberato}
\affiliation{School of Physics and Astronomy, University of Southampton, Southampton, SO17 1BJ, United Kingdom}

\begin{abstract}
We experimentally investigate second harmonic generation from strongly coupled localized and propagative phonon polariton modes in arrays of silicon carbide nanopillars. Our results clearly demonstrate the hybrid nature of the system's eigenmodes and distinct manifestation of strong coupling in the linear and nonlinear response. While in linear reflectivity the intensity of the two strongly-coupled branches is essentially symmetric and well explained by their respective localized or propagative components, the second harmonic signal presents a strong asymmetry. Analyzing it in detail, we reveal the importance of interference effects between the nonlinear polarization terms originating in the bulk and in the phonon polariton modes, respectively.
\end{abstract}


\maketitle


Controlled confinement of light in sub-diffraction volumes has always been a key goal in photonics. In this regard, plasmonic systems provide rich opportunities for manipulation of light on the nanoscale\cite{MaierBook,OdomCR11,BrongersmaNatNano15}. Recently, an alternative approach utilizing polar dielectrics operating in the mid-infrared (mid-IR) spectral range has attracted considerable attention \cite{CaldwellNanoLett13,DaiScience14,CaldwellNatMater15,YoxallNatPhot15,GubbinPRL16,LiNatMater16,GubbinACSPhot17a,GubbinACSPhot17b,PasslerJOSAB17,Dunkelberger2018,BerteACSPhot2018}. 
The photonic modes of these systems, termed surface phonon-polaritons, exhibit low optical losses due to their relatively large phonon lifetimes, allowing for higher degrees of energy concentration and smaller Purcell factors than their plasmonic counterparts \cite{KhurginNanophotonics17}.
Conventional methods of probing the field enhancement include multi-photon effects such as surface enhanced Raman scattering, optical rectification, and second harmonic generation (SHG). 

Bearing strong conceptual similarities to nonlinear plasmonics in metals \cite{KauranenZayats12}, efficient SHG mediated by localized and propagating phonon-polariton modes in polar dielectrics has been reported \cite{AlievaJETP95,DekorsyPRL03,RazdolskiNanoLett16,PasslerACSPhot17}. 
In contrast to noble metals supporting surface plasmon-polaritons, the lack of inversion symmetry enables bulk SHG in these systems, opening intriguing questions about the nonlinear-optical response of phonon-polaritons.
In 2016 Gubbin and coworkers observed strong coupling between phonon-polariton modes localised in SiC nanopillars and propagative modes sustained on the substrate surface, giving rise to the two hybrid localised-propagative polariton branches \cite{GubbinPRL16}.
While the mechanisms of SHG enhancement in plasmonic and phononic structures are reasonably well understood, the nonlinearity of coupled systems where multiple interacting resonances hybridize remains widely unexplored. Previous works on coupled localized and propagating plasmon modes \cite{HollandPRL84,BorenszteinPRL84,StuartPRL88,NordlanderNanoLett04,ChristPRB06,GhoshalAPL09,SarkarACSPhot15} focused on the linear optical response and its tunability prospects, meaning that little is known about the nonlinear-optical properties of the coupled polaritonic modes. 

In this Article, we study the nonlinear-optical response of coupled surface phonon-polaritons in arrays of SiC nanopillars. Employing a free-electron laser (FEL) as a powerful, tunable source of mid-IR radiation we perform SHG spectroscopy on a series of samples with varied array pitch,
thus tuning the surface phonon-polariton resonance.
Our results demonstrate pronounced differences between the spectral reflectivity and SHG response at the avoided crossing in the polariton dispersion. We further outline relevant mechanisms for the observed SHG spectral behaviour and discuss the key role of the substrate for the disparate optical response of the coupled modes in the linear and nonlinear domains. 

We perform spectroscopic measurements in the IR range, recording both linear reflectivity and SHG from square arrays of SiC nanopillars ($0.8~\mu$m high, $1~\mu$m in diameter) etched on a 3C-SiC substrate. The fabrication procedure has previously been described in the literature \cite{GubbinPRL16}.
The details of the FEL are also reported elsewhere \cite{SchollkopfFEL}. Here it is tuned through a wavelength range of $9-13$ $\mu$m (frequency $750-1050$ cm$^{-1}$) and impinges on the sample at an angle of incidence of $\theta=62^{\circ}$ from the surface normal. The reflected fundamental radiation and SHG output were registered by a home-built pyroelectric photodetector and a liquid nitrogen-cooled mercury cadmium telluride
detector (Infrared Associates), respectively.

\begin{figure}
    \includegraphics[width=0.9\columnwidth]{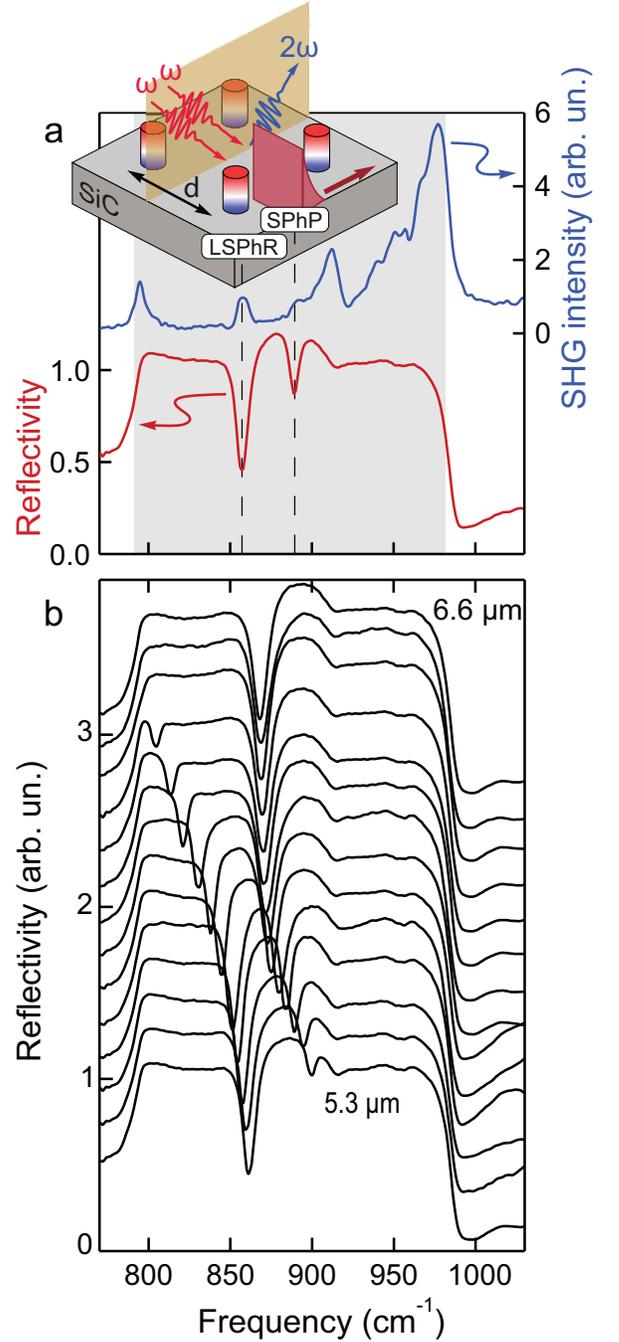}
    \caption{Strong coupling of a localized surface phonon resonance (LSPhR) and a propagating surface phonon-polariton (SPhP) mode. (a) Typical linear reflectivity (red) and SHG intensity (blue) spectra measured on a nanopillar array with pitch $d=5.5~\mu$m. The vertical dashed lines indicate the positions of the surface phonon resonances. The gray box shows the SiC Reststrahlen band. Inset: sketch of the experimental geometry and the excited propagating and monopole localized polariton modes. (b) Linear reflectivity plots measured on a series of arrays showing the shift of the resonances with varied pitch due to the localised-propagative mode coupling.
    }
    \label{fig:typical}
\end{figure}

Localized phonon polaritons are supported in the individual nanopillars \cite{CaldwellNanoLett13,ChenACSPhot14,RazdolskiNanoLett16}, while propagative surface phonon polaritons are supported by the SiC substrate \cite{HuberAPL05,PasslerACSPhot17,PasslerNanoLett18}.
The periodicity of the nanopillars enables phase-matched coupling of far field, p-polarised radiation into the propagative modes \cite{LeGallPRB97,GreffetNature02} which can hybridize with the localized modes of the pillars \cite{GubbinPRL16,GubbinPRB17}. These modes are observed as narrow dips on the high Reststrahlen reflectivity background of the substrate, as well as pronounced peaks in the SHG spectra due to strong light localization and enhancement of the electromagnetic field. Typical spectra shown in Fig.~\ref{fig:typical}a were obtained on a sample with the array pitch $d = 5.5~\mu$m. The positions of the features associated with the excitation of surface phonon resonances are indicated with the dashed vertical lines.
Note that at this particular array pitch, the coupling is weak, allowing the observation of the two fundamental modes in an almost unperturbed regime. The monopole mode appears at about $865$~cm$^{-1}$ while the frequency of the propagative phonon-polariton (here $\approx890$~cm$^{-1}$) shifts with the array pitch.
In what follows, we shall focus on those features and do not discuss the SHG peaks at the frequencies of the transverse ($\approx 790$~cm$^{-1}$) and longitudinal ($\approx 980$~cm$^{-1}$) optical phonons in SiC which are the fingerprints of the SiC crystalline symmetry. These features have been analyzed previously \cite{PaarmannAPL15,PaarmannPRB16} and are insensitive to the pitch $d$ of the nanopillar array.

\begin{figure}
    \includegraphics[width=0.95\columnwidth]{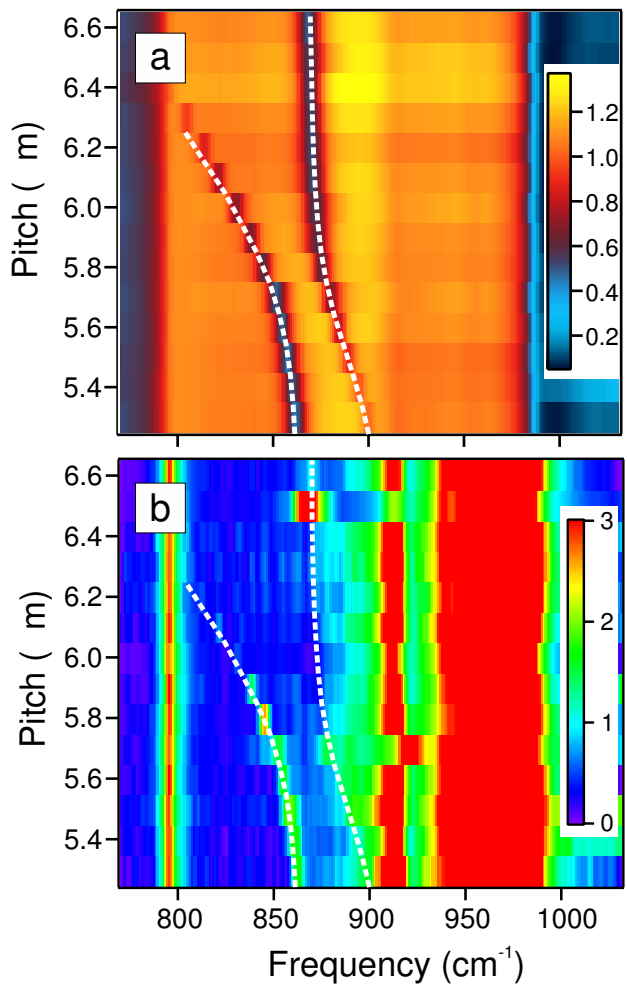}
    \caption{False color experimental spatio-spectral maps of linear reflectivity (a) and SHG intensity (b) measured on a set of arrays of nanopillars with varied pitches $d$. The white dashed lines illustrate the calculated dispersion of the strongly coupled polariton modes.}
    \label{fig:maps}
\end{figure}

Strong coupling of the polaritonic modes can be observed if the array pitch enables the excitation of propagating surface phonon polaritons at a frequency near to that of a localized mode \cite{GubbinPRL16}. In particular, this results in significant hybridization of the propagating phonon-polariton with a so-called monopolar mode \cite{ChenACSPhot14,CaldwellNanoLett13,GubbinPRB17} accompanied with the avoided crossing behaviour. To study the strong coupling regime in detail and map out the dispersion of the coupled modes in the vicinity of the avoided crossing, we performed spectroscopic measurements on a series of arrays with varied pitch $d$ in the range of $5.3 - 6.6~\mu$m. The reflectivity data (Fig.~\ref{fig:typical}b) are in a good agreement with the previously published results \cite{GubbinPRL16,GubbinPRB17}, demonstrating the avoided crossing of the dispersion curves of the two phonon-polariton modes. As in this work we are mostly interested in the nonlinear-optical response of the coupled polaritonic modes, we compare the SHG spatio-spectral map with that in the linear domain. The direct comparison (Fig.~\ref{fig:maps}) reveals similar behaviour of the dispersion of the coupled modes. 
Further, the coupling of the propagating surface mode to the monopolar localized phonon-polariton results in noticeable spectral shifts of the linear extinction and SHG output peaks, demonstrating that the nonlinear emission originates in the hybrid localized-propagative modes.

\begin{figure}[t]
    \includegraphics[width=0.85\columnwidth]{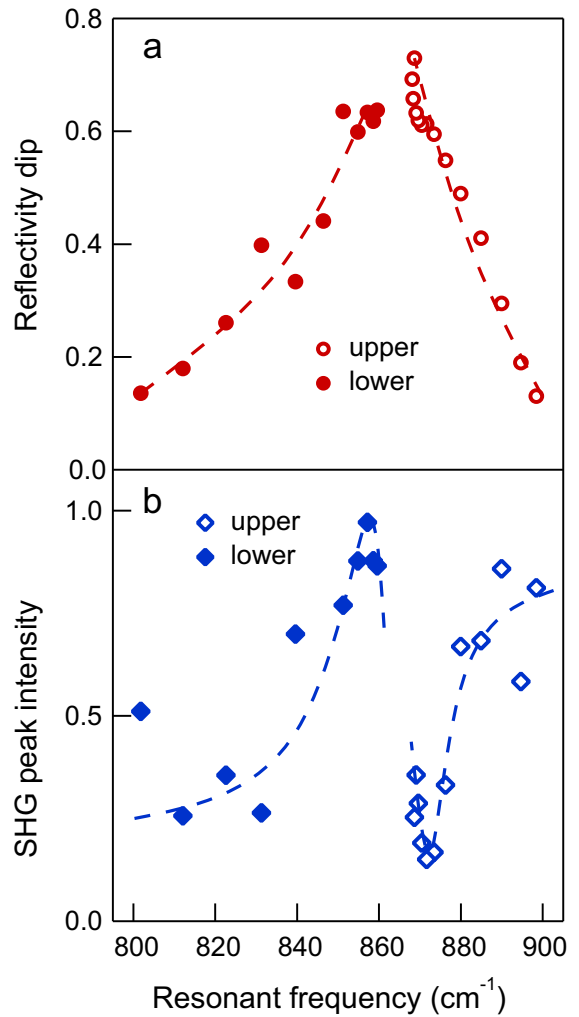}
    \caption{Analysis of the coupled polariton modes. Magnitudes of the reflectivity dip (a) and peak SHG intensity (b) along the dispersion branches of the strongly coupled polariton modes. The dashed lines are guides to the eye.}
    \label{fig:peaks}
\end{figure}

Notably, in Fig.~\ref{fig:maps}b it is clearly seen that in the SHG output the low-frequency polariton branch is more strongly pronounced than the high-frequency one, while in the reflectivity map the two branches are almost symmetric. To better understand the difference in the linear reflectivity and SHG output spectra we analysed the magnitudes of the SHG peaks and the reflectivity dips along the dispersion curves modified by the strong coupling. In order to do this, for each pitch $d$ we extracted the reflectivity and SHG data at the frequencies given by the dispersion of the coupled modes $\omega^{\pm}$:

\begin{equation}
    \label{coupled}
    \omega^{\pm}=\frac{\omega_1+\omega_2\pm\sqrt{(\omega_1-\omega_2)^2+4g^2}}{2},
\end{equation}
where $\omega_{1,2}$ are the frequencies of the localized and propagative modes, and $g$ is the coupling strength (Rabi frequency). The results of this procedure are shown in Fig.~\ref{fig:peaks}, where the open and full symbols refer to the upper ($\omega^{+}$) and lower ($\omega^{-}$) polariton branches, respectively.

In Fig.~\ref{fig:peaks}a we plot the depth of the corresponding dip in the reflectivity, that is, the difference between the high reflectance within the Reststrahlen band observed on a flat SiC surface and the respective data point obtained on the array of nanopillars. Each pair of points (consisting of an open and a closed symbol) in Fig.~\ref{fig:peaks} corresponds to a particular array pitch $d$. The pronounced increase of the depth upon approaching the frequency of the monopolar mode $\omega_2\approx~865$~cm$^{-1}$ is related to the hybrid character of the two coupled modes. While the monopolar localized mode couples strongly to the far-field radiation, suboptimal grating conditions mean that the excitation of the surface phonon-polariton with free-space photons is inefficient.
This inequality results in deeper reflectivity dips when a hybrid mode is predominantly monopolar in character, as it can be seen in Fig.~\ref{fig:typical}. The reflectivity dip depth of the monopolar mode (around $865$~cm$^{-1}$) is significantly larger than the depth at the propagative mode far away from it. 
When the detuning of the bare, uncoupled polaritons approaches zero, 
their hybridization results in the coupled modes
comprised of approximately equal proportions of monopolar localized and propagative modes, leading to the equilibration of the depths of the resonances. The slight asymmetry visible in the reflectivity data (Fig.~\ref{fig:peaks}a) can be attributed to the upper polariton (at frequency $\omega^+$) engaging in a quasi-resonance with the localized transverse dipolar mode at around  $912$~cm$^{-1}$. The resulting avoided coupling slightly red-shifts the upper polariton branch, leading to the small difference in the slopes for the two polaritonic modes.

However, the hybridization between the two disparate polariton modes upon their strong coupling
alone cannot account for the striking asymmetry in the SHG intensity from the coupled modes. It is seen in Fig.~\ref{fig:peaks}b that the SHG output along the lower polariton branch increases upon approaching the monopolar mode from the low frequency end, whereas the upper branch demonstrates the opposite trend.
The larger SHG output of the propagative mode on the high-frequency end of the spectrum ($\approx 900$~cm$^{-1}$) compared to that on the opposite side ($\approx 820$~cm$^{-1}$) originates in the interplay of the dispersion of the SiC nonlinear susceptibility $\chi^{(2)}$ and the effective Fresnel factors for the electric field including its polariton-driven enhancement \cite{PaarmannPRB16,GubbinACSPhot17a}. The former factor peaks at the transverse optical phonon resonance at about $780$~cm$^{-1}$, whereas the latter is responsible for the SHG peak at the longitudinal optical phonon frequency ($\approx 980$~cm$^{-1}$). The characteristic SHG spectrum shown in Fig.~\ref{fig:typical}a, consistent with previous results of SiC SHG spectroscopy \cite{PaarmannAPL15,PaarmannPRB16}, indicates the higher importance of the electric field enhancement factor, giving rise to the steadily increasing SHG background at larger frequencies in the spectral range of $830-950$~cm$^{-1}$.

\begin{figure}
    \includegraphics[width=0.9\columnwidth]{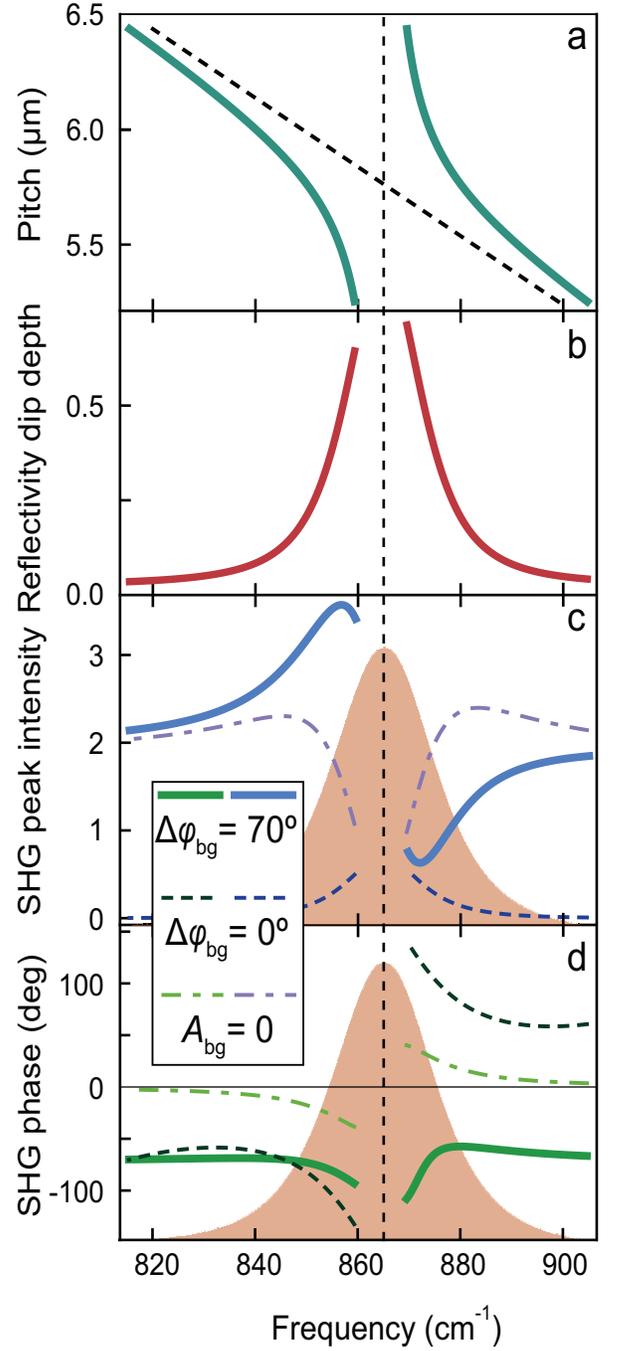}
    \caption{Two coupled oscillators model. (a) Dispersion of the uncoupled (dashed black lines) and the strongly coupled modes (solid green lines). (b,c) Simulated reflectivity dip (b, solid red lines) and SHG peak intensity (c, solid blue lines) along the branches of the coupled modes. The shaded area in (c) illustrates the lineshape of a localized oscillator. (d) The simulated phase of the nonlinear polarization $P^{2\omega}$ in the vicinity of the two resonances. The solid lines in (c-d) were calculated with a phase of the complex background contribution $A_{\rm bg}$ $\varphi_{\rm bg}=-70^{\circ}$ with respect to that of the resonant Lorentzians (Eq.~\ref{lorentz}), resembling the experimental observations. Such agreement cannot be met when $A_{\rm bg}$ is set to have the same phase as $L_k$, or $A_{\rm bg}=0$, as is illustrated by the dashed and dot-dashed lines, respectively.
    }
    \label{fig:model}
\end{figure}

Summarizing these observations, we note that the SHG spectrum of the coupled polaritons demonstrates strikingly disparate behaviour to the relatively simple and intuitive picture obtained in the linear reflectivity measurements. To understand this difference, we employ a simple model of two coupled oscillators giving rise to the SHG signal on top of the background contribution. For clarity, we neglect the weak interaction of the propagative polariton with the aforementioned transverse dipolar localized mode. Within this model, the resonances can be described by the two Lorentzians:

\begin{equation}
    \label{lorentz}
    L_k(\omega)=\frac{A_k}{\omega-\omega_k+i\gamma_k}e^{i\varphi_k},
\end{equation}
where  $k=1,2$, $\omega_k$ and $\gamma_k$ are the resonant frequencies and damping of the two modes, and $\varphi$ indicates a possible phase shift. In our model, $\omega_1$ is set constant to $865$~cm$^{-1}$ (to match the monopole mode frequency) while $\omega_2$ can be tuned by an external parameter, i.e., the array pitch $d$, representing the dispersion of the propagative mode and its grating-mediated excitation. To simulate the propagative-monopole mode coupling, we continuously change the control parameter $d$ so that $\omega_2$ sweeps across $\omega_1$, and the resonant frequencies of the coupled modes are given by Eq.~\ref{coupled} (see Fig.~\ref{fig:model}a). There, the extinction of the electric field $E(\omega)$ is given by the imaginary part of the linear-optical response at the frequencies $\omega^{\pm}$:

\begin{equation}
    \label{absorption}
    r(\omega^{\pm})\propto\mathrm{Im}\left[ L_1(\omega^{\pm})+L_2(\omega^{\pm}) \right],
\end{equation}
where the permutation of $\omega^{\pm}$ does not change the result. In turn, the depth in the experimental reflectivity data is proportional to $\lvert r(\omega^{\pm})\rvert^2$ (Fig.~\ref{fig:model}b). Here we neglect the absorption in the SiC substrate due to its high reflectivity within the Reststrahlen band.

On the contrary, the SHG response needs to include the background contribution of the substrate. We can write down the nonlinear polarization at the double frequency $P^{2\omega}$ created by the excitation at the fundamental frequency $\omega$ in the following way:

\begin{equation}
    \label{shg}
    P^{2\omega}(\omega^{\pm})\propto \chi^{(2)}_1L_1^2(\omega^{\pm})+\chi^{(2)}_2L_2^2(\omega^{\pm})+A_{\rm bg}.
\end{equation}
Here the background contribution $A_{\rm bg}$ can be complex, and $\chi^{(2)}_k$ are the effective nonlinear-optical susceptibilities of the two modes. The resonant increase of the SHG efficiency is attributed to the
the polariton-induced enhancement of the electric field $E\propto L_k$.   
%
For instance,
the SHG enhancement at the localized monopolar mode originates in the subdiffractional localization of the electric field and the prominent increase of its out-of-plane component $E_z$ \cite{RazdolskiNanoLett16,GubbinACSPhot17a}. The surface-mode-induced SHG is additionally quenched by the low efficiency of the grating consisting of the nanopillars, as discussed above. Within our model, the experimentally observed SHG intensities at the frequencies of the coupled modes are given by $\lvert P^{2\omega}(\omega^{\pm})\rvert^2$.

The results of our modeling are summarized in Fig.~\ref{fig:model}, nicely resembling our experimental observations. The slight asymmetry between the two branches in the reflectivity (Fig.~\ref{fig:model}b) is introduced by the phase shift between the localized and propagative resonances, $\Delta\varphi=\varphi_2-\varphi_1$. We found that the degree of asymmetry commensurate with the experimental observations can be obtained for $\lvert\Delta\varphi\rvert<5^{\circ}$. 
Taking into account the simplicity of our model and possible quasi-resonance with the transverse dipolar polariton, we can conclude on the zero phase shift between the two considered polariton modes. 
At the same time, the phase shift between the resonant and the background SHG contributions appears crucial for the asymmetry in the SHG dependence. The antisymmetric shape seen in Fig.~\ref{fig:model}c can only be obtained when the phase shift of the complex background contribution $A_{\rm bg}$ with respect to the resonant Lorentzians $\lvert\Delta\varphi_{\rm bg}\rvert=\lvert\varphi_{\rm bg}-\varphi_{1,2}\rvert$ is set to about $90^{\circ}$.
In particular, the results of our simulations shown with the solid lines in Fig.~\ref{fig:model}c were obtained for $\Delta\varphi_{\rm bg}\approx70^{\circ}$.

To demonstrate the importance of the phase-shifted background contribution to the total SHG output in the vicinity of the strong coupling, in Fig.~\ref{fig:model}c we further plot the shapes of the resonant SHG yields if the phase shift $\Delta\varphi_{\rm bg}$ vanishes (dashed lines) and if the background contribution is not taken into account at all (dash-dotted lines). It is seen that the experimentally observed degree of asymmetry in the resonant SHG response can only be obtained in simulations if the background contribution $A_{\rm bg}$ with its own distinct phase is considered. If either $\varphi_{\rm bg}$ or the entire $A_{\rm bg}$ is set to zero, it is further seen that when $\lvert\Delta\varphi\rvert\approx0$ (as enforced by the linear reflectivity data), 
the asymmetry of the two branches of the coupled resonances is eventually lost,
thus ruling out the reproduction of the experimentally achieved SHG behaviour. Fig.~\ref{fig:model}d provides additional insight into the nonlinear optics of the strongly coupled resonances, showing the simulated phases of the nonlinear polarization $P^{2\omega}$ in these three cases.   
The curves in Fig.~\ref{fig:model} were simulated for a coupling strength of $g=15$~cm$^{-1}$, which is in a good agreement with the values found from the analysis of the experimental data \cite{GubbinPRL16}.

As such, this disparate behaviour of the resonant linear and nonlinear properties is an interesting optical phenomenon originating in the utter complexity of the nonlinear-optical response \cite{UtikalPRL11,ChekhovPRB16}.
We emphasize that the goal of our modelling was not to accurately reproduce the experimental data (which would require inclusion of too many fitting parameters in the model) but rather to demonstrate how the main features of the SHG intensity variations in the vicinity of the coupled resonances, which exhibit drastically different behaviour to that of the linear reflectivity data, can be relatively simply understood. With the help of our modelling we found that the interference of the resonantly enhanced SHG waves with the phase-shifted background contribution produced by the substrate is a key for the asymmetry between the two branches of the coupled polaritonic modes. Importantly, this phase shift is inherent to the polaritonic nature of the modes: on top of the well-known enhancement of the electric field amplitude, the surface polaritons (both localized and propagating) are characterized by the phases of various electric field components. 
In other words, the phase shift of the electric fields associated with the particular surface polariton mode is then imprinted into the phase of the corresponding nonlinear polarization term $P^{2\omega}$.
Because our model does not explicitly consider the polariton-driven electric field enhancement, this effect is instead taken into account by introducing the effective phases $\varphi_k$ and susceptibilities $\chi^{(2)}_k$. As such, not only the amplitude but also the phase of the resulting total $P^{2\omega}$ can vary between the resonant and non-resonant cases, giving rise to the interesting interference conditions at the strong coupling resonance.

To summarize, we have analyzed the SHG response of strongly coupled surface phonon-polariton modes. In particular, we observe SHG from the normal modes of an array of SiC nanoresonators, consisting of hybridized localized and propagating surface phonon-polaritons in the SiC Reststrahlen band. The far-field excitation of the polaritons is enabled by the periodicity of SiC nanopillar arrays with a systematically varied pitch. In contrast to the linear reflectivity measurements, we found a clear antisymmetric behaviour of the resonant SHG output along the two dispersion branches of the coupled polaritons. Employing a simple coupled oscillator model, we demonstrate that the disparate symmetry of the linear and SHG response can be explained by the interference of the polariton-induced SHG with the background contribution. We further argue that the polaritonic enhancement of the electric fields and their phase shifts are responsible for the particular interference conditions leading to the experimentally observed asymmetry. Our results advance the understanding of the nonlinear nanophononics in the mid-infrared spectral range, while retaining a high degree of generality and thus remaining valid for e.g. surface plasmon-polaritons in metals.

\begin{acknowledgements}
The authors thank W. Sch{\"o}llkopf and S. Gewinner for their assistance with operating the free electron laser. S.D.L. is a Royal Society Research Fellow. S.D.L and C.R.G. acknowledge support from EPSRC Grant No. EP/M003183/1. S.A.M. acknowledges the Lee-Lucas Chair in Physics and the Leverhulme Trust.
\end{acknowledgements}


\bibliography{refs}

\end{document}